# Towards Runtime Customizable Trusted Execution Environment on FPGA-SoC

Yanling Wang, Xiaolin Chang, Haoran Zhu, Jianhua Wang, Yanwei Gong and Lin Li

**Abstract**—Processing sensitive data and deploying well-designed Intellectual Property (IP) cores on remote Field Programmable Gate Array (FPGA) are prone to private data leakage and IP theft. One effective solution is constructing Trusted Execution Environment (TEE) on FPGA-SoCs (FPGA System on Chips). Researchers have integrated this type TEE with Trusted Platform Module (TPM)-based trusted boot, denoted as FPGA-SoC tbTEE. But there is no effort on secure and trusted runtime customization of FPGA-SoC TEE.

This paper extends FPGA-SoC tbTEE to build Runtime Customizable TEE (RCTEE) on FPGA-SoC by additive three major components (our work): 1) CrloadIP, which can load an IP core at runtime such that RCTEE can be adjusted dynamically and securely; 2) CexecIP, which can not only execute an IP core without modifying the operating system of FPGA-SoC TEE, but also prevent insider attacks from executing IPs deployed in RCTEE; 3) CremoAT, which can provide the newly measured RCTEE state and establish a secure and trusted communication path between remote verifiers and RCTEE. We conduct a security analysis of RCTEE and its performance evaluation on Xilinx Zynq UltraScale+ XCZU15EG 2FFVB1156 MPSoC.

**Index Terms**—Field Programmable Gate Array, Intellectual Property, Remote Attestation, System on Chip, Trusted Execution Environment.

## 1 Introduction

Large-scale data processing in cloud data centers advances the widespread deployment of hardware accelerators like Field Programmable Gate Array (FPGA) [1], Application Specific Integrated Circuit (ASIC) [2] and Graphic Processor Unit (GPU) [3]. Benefiting from its dynamic re-configurability, FPGA performs more flexibly and gains huge interests as an acceleration solution [1][4]. Cloud service providers, such as Microsoft, Alibaba, Amazon and Huawei, are actively adopting FPGA with unmatched capabilities and intrinsic advantages into their infrastructures [5]. However, in a cloud FPGA, sensitive data and well-designed Intellectual Properties (IPs) of users are prone to privacy data leakage and IP theft [6].

Trusted Execution Environment (TEE) addresses these privacy and security issues by providing a secure processing environment, namely, isolating the critical data and applications of users from that in non-secure processing environment (referred to as Rich Execution Environment, REE) [7][8]. TEE is originally designed for Central Processing Unit (CPU) but this purely CPU-based TEE has efficiency issue [9]. Researchers have explored the implementation of TEE on FPGA (denoted as FPGA-TEE) and existing solutions can be classified into the following three categories, illustrated in Fig.1.

(Solution_a) **Pure FPGA TEE**. Use the whole FPGA as an independent TEE [10]-[13], shown in Fig.1.a. This type solutions divide TEE into static logic and dynamic logic on FPGA. Remote users communicate with the static logic, which is responsible for deploying IPs in the region of dynamic logic.

(Solution_b) **Host-FPGA TEE**. Use the whole FPGA as an extension of CPU-TEE [14], shown in Fig.1.b. The Host-FPGA TEE system is formed by extending CPU-based TEE on a host to FPGA. The authors in [14] and [15] explored ARM TrustZone and Intel SGX for Host-FPGA TEE, respectively.

(Solution_c) **FPGA-SoC TEE**. Use part of FPGA-SoC (FPGA System on Chip) as FPGA TEE and the left as REE [17], shown in Fig.1.c. ARM TrustZone as processor-built-in security technology has been implemented on FPGA-SoC [16][18][19] for improving FPGA-SoC security. According to whether leveraging processor-built-in security technology, this type solutions are further divided into two sub-categories: (c.1) no processor-built-in security technology is used [21]-[23] and (c.2) processor-built-in security technology is used [17][20]. Note that both these two sub-categories have the same structure shown in Fig.1.c.

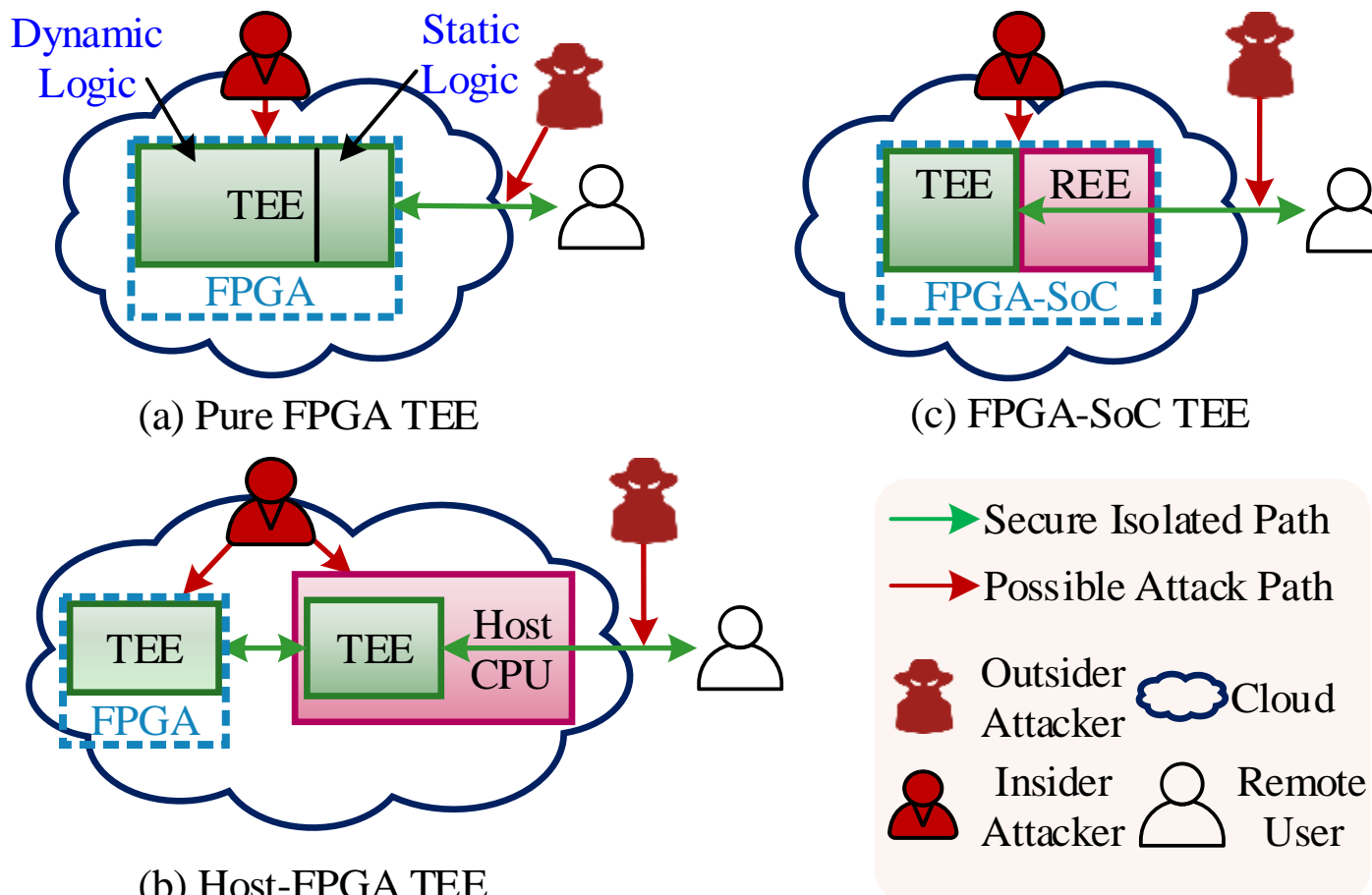

Fig. 1. FPGA-TEE Solutions.

There are at least three issues with Solution_a: (P1) A network server must run inside TEE for handling remote user requests, which requires an open service port and then increases the attack surface of FPGA-TEE. (P2) Static logic occupies part of hardware acceleration resources at all time even they are not used, which reduces the effective utilization of FPGA resources. (P3) Since static logic and dynamic logic are not isolated from each other, a compromised or malicious static logic can easily cause threats to dynamic logic.

There are at least two issues with Solution_b: (P1) Only the software cryptographic algorithms and pseudo random seed generators implemented on CPU-TEE can be used. Consequently, it is hard, if not impossible, to dynamically replace software cryptographic algorithms. (P2) FPGA is susceptible to all security attacks from CPU-TEE.

For Solution_c.1, the execution space of FPGA-TEE is not separated from that of REE, which enables inside attackers to control part of FPGA-TEE through the software and/or hardware on REE exposed to them.

Solution_c.2 can address the above-mentioned problems, motivating our work. This paper focuses on a flexible and secure provision of FPGA-SoC TEE. Khan et.al. [17] developed a FPGA-SoC TEE by utilizing ARM TrustZone, named as Open Portable Trusted Execution Environment (OP-TEE). However, OP-TEE does not provide trusted boot (referred to as measured boot proposed in [17]). To this end, Gross et.al. [20] provided a FPGA-SoC TEE with trusted boot (denoted as **FPGA-SoC tbTEE** in this paper), where each of the components (including OP-TEE) involved in the booting process of FPGA-SoC are measured via hash values. This enables remote attestation. However, FPGA-SoC tbTEE is a static TEE and then it does not allow the customization of FPGA-SoC TEE at runtime. Note that the works in both [17][20] can deploy IPs in REE at runtime, which is not secure.

In this paper, we aim to construct a *R*untime *C*ustomizable FPGA-SoC *TEE* (**RCTEE**),by extending the FPGA-SoC tbTEE. The extension includes three major components: CrloadIP, CexecIP and CremoAT. To the best of our knowledge, it is the first time to investigate customizable FPGA-SoC TEE at runtime. Consequently, RCTEE (illustrated in Fig.4) not only owns the security functionalities of **FPGA-SoC tbTEE** but also has the following unique functionalities:

1) **Secure runtime customization of RCTEE.** CrloadIP developed in RCTEE handles the secure dynamic IP deployment at runtime. Unless an IP passes the integrity checking by CrloadIP, it can not be deployed in FPGA-SoC TEE.
2) **Secure and flexible usage of IPs in RCTEE.** CexecIP developed in RCTEE is responsible for this functionality. A unified input and output format for IPs is provided in CexecIP, therefore any IP deployed at runtime can be used without modifying operating system of FPGA-SoC TEE.
3) **Secure and trusted data transmission between RCTEE and remote users.** CremoAT is responsible for providing the newly measured RCTEE state. In addition, CremoAT provides a remote attestation protocol to allow the remote user to verify the trustworthiness of newly booted FPGA-SoC. A secure and trusted communication path can be established between remote verifiers/users and RCTEE without increasing the attack surface of FPGA-SoC TEE, addressing the first security issue (P1) of Solution_a. It is done by implementing a proxy server in REE for forwarding data between RCTEE and remote users. Only RCTEE and user can decrypt the encrypted data and IP cores in bitstream form.

Our security analysis indicates that RCTEE can resist cryptographic attacks, network attacks including replay and Man-in-the-middle attacks, and physical attacks from inside attackers. The performance evaluation on Xilinx Zynq UltraScale+ XCZU15EG 2FFVB1156 MPSoC indicates our approach's practicability.

The remainder of the paper is set up as follows: Section 2 and Section 3 introduce background and related works, respectively. Section 4 describes the system considered in this work and an overview of the proposed RCTEE approach. We present the implement details in Section 5 followed by security and performance analysis about this work in Section 6. Section 7 concludes this paper.

## 2 BACKGROUND

This section presents the background for understanding the related components and concepts in this paper.

### 2.1 FPGA-SoC of ZYNQ Ultrascale+ Platform

The Xilinx Zynq Ultrascale+ MPSoC family is based on the Zynq Ultrascale+ (ZU+) architecture. As presented in Fig.2, this family of FPGA-SoCs integrates programmable logic (PL) and a feature-rich processing system (PS) on a single device. The PL consists of configurable fabric (FPGA fabric) of which the functionality is determined by a configuration memory. Data stored in the configuration memory is referred to as the bitstream or configuration data. The units of PS used in our work include an ARM Cortex-A53 Application Processor Unit (APU), a Platform Management Unit (PMU) and a Configuration Security Unit (CSU) [18]. In this paper, we leverage CSU to achieve secure and trusted boot of the entire FPGA-SoC, including PS and PL. CSU incorporates hardware cryptographic accelerators (AES256, SHA3-384 and RSA4096/1024), key management unit indispensable to secure boot, a built-in DMA (CSUDMA) and a PCAP interface used for loading bitstream to PL via software. In short, CSU is responsible for booting PS in a secure or non-secure mode and also for configuring PL. Besides a large external Double Data Rate (DDR) memory, the FPGA-SoC contains 256 KB on-chip memory (OCM) which can be used for storing sensitive data or codes [24].

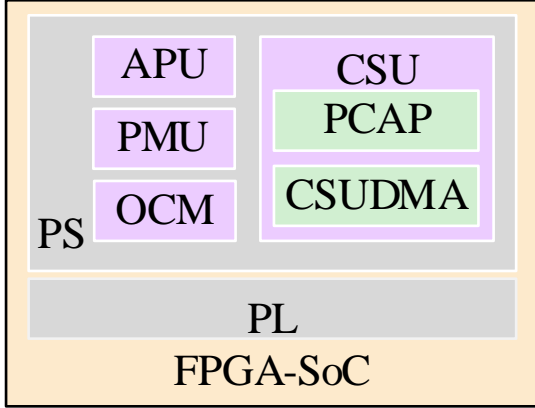

Fig. 2. FPGA-SoC of ZU+ architecture.

### 2.2 Secure Boot Mechanism

The ZU+ devices support secure boot mechanisms, with optional authentication, encryption or validation engines to protect boot files for PS and configuration data for PL. There exist two secure boot modes: (1) hardware root of trust mode and (2) encrypt only mode [25]. The former uses several RSA keypairs programmed on eFuse (a fixed and non-update-able memory) to provide the authentication of boot files and bitstream, with optional encryption and integrity-checking, to provide confidentiality and integrity. The latter is a boot mechanism that utilizes an AES-GCM engine and a key programmed on eFuse to decrypt a bootable image generated by encrypting and packaging all the files [25]. AES keys used for hardware root of trust mode can be programmed in eFuse or battery backed RAM (BBRAM). Note that keys programmed in eFuse or BBRAM for secure boot are simplified as device keys in this paper.

### 2.3 Physical Unclonable Function

Physical Unclonable Function (PUF) is a hardware security technology which uses the manufacturing variation of a silicon nanocircuit to identify these devices and derive entropy sources [27]. It is always used for device authentication to prevent FPGA impersonation attack [11], and provide random seeds for key generation. Our paper adopts PUF to both authenticate FPGA-SoC and obtain random seeds for generating keypairs.

### 2.4 ARM TrustZone and OP-TEE

ARM TrustZone can be utilized to building TEE in PS and also can extend the security of TEE in PS to PL, which achieves TEE across PS and PL.

**TEE in PS**. ARM TrustZone for Cortex-A processors inside FPGA PS provides a system-wide isolation between a secure world (SeW) and a non-secure world (non-SeW), which is achieved by partitioning both hardware and software resources. ARM Trusted Firmware (ATF) contains a Secure Monitor for managing the switching and communication between two worlds. OP-TEE is a TEE designed in [28] as a companion to a non-secure Linux kernel running on ARM Cortex-A cores using the ARM TrustZone technology. OP-TEE implements TEE Internal Core API and TEE Client API. Applications in REE (non-SeW) can interact with trusted applications (TAs) in TEE (SeW) via Secure Monitor, which is accessible from the user space via TEE Client APIs. A TA running in TEE can access the trusted functionalities provided by TEE only through the TEE Internal Core API [28][29].

**Security extension to PL**. The security of TEE can be extensible to the FPGA PL by setting the security bit in an AXI Interconnect IP. A master interface of AXI Interconnect can declare the AXI peripheral connected to its slave interface as a secure or non-secure IP. Two internal bits of the AXI peripheral, namely AWPROT for write transaction and ARPROT for read transaction, inform the full system that the executed software is running in TEE or REE. Only TAs running in OP-TEE can access a secure IP, while a non-secure IP is accessible by applications running in TEE and REE [24][30][31].

### 2.5 Threat Model and Security Requirements

The most valuable assets in a cloud FPGA-SoC contains the private data and IPs developed by cloud service users. Appropriate measures must be taken to ensure that those assets are protected against inside and outside attacks [11]. There are at least the following three types of threats to FPGA-SoC TEE.

1) **Obtain sensitive data or IPs of remote users during transmission between FPGA-SoC TEE and the user**. Attackers may attempt to obtain sensitive data or theft IPs by deceiving the user with an impersonated device or breaking the cryptographic security of session key shared between FPGA-SoC TEE and the user.
2) **Inject malicious IPs**. Attackers inside or outside the cloud attempt to inject a malicious IP into PL, trigger it, and then invoke, theft or tamper with the IPs or data created by users.
3) **Make unauthorized access to resources in TEE.** Inside attackers have physical access to the cloud device and is able to control part of privileged software and hardware through the REE exposed to cloud employees. Inside attackers can read out IPs in bitstream form via reconfiguration interfaces even when the bitstream is encrypted before the device being reconfigured [17]. They can also perform unauthorized and arbitrary invocation of IPs created by users. It is also possible that inside attackers attempt to obtain the device keys and reverse engineering the encrypted bootable image.

To defend against the threats above, RCTEE at least needs to meet the following security requirements (**SR**).

(1) **SR1: TEE Verifiability**. The cloud device is booted with encrypted bootable image involving integrity checking, which ensures bitstream, firmware and software are measured before loading into the device. Note that the measurements of every component in bootable image should be stored securely for later remote attestation.

(2) **SR2: FPGA-SoC Authentication.** FPGA-SoC authentication ensures users that the device in the cloud is a specific physical FPGA-SoC instead of an emulated hardware or software module.

(3) **SR3: Secure and Trusted TEE Runtime Customizing.** If the security of TEE runtime customizing or the integrity of bitstream used to customize TEE is not guaranteed, TEE can easily be compromised.

(4) **SR4: Secure IP Invocation.** A secure IP invocation interface should be developed and integrated in OP-TEE. Only TEE can invoke IPs and cloud employees cannot invoke IPs via REE exposed to them.

(5) **SR5: Secure Transmission.** No plain text data or bitstream is accessible to entities other than FPGA-SoC TEE and the user who creates these assets. If any private keys for remote attestation of FPGA-SoC TEE and the users are exposed to any third party (TTP) involved, TTP can easily decode the IPs and data encrypted with a session key shared between FPGA-SoC TEE and remote users.

## 3 Related Work

There are researches on developing TEEs for protecting user data and computation on FPGA. Pure FPGA TEE was developed by authors in [10]-[13], and the authors in [14] and [15] explored the establishment of Host-FPGA TEE. The rest of this section focuses on investigating the approaches to building FPGA-SoC TEE. Note that ARM TrustZone is the only processor-built-in security technology used in FPGA-SoC for building FPGA-SoC TEE so far. ARM TrustZone can prevent inside attackers from control part of FPGA-TEE through the software and/or hardware on REE exposed to them. Therefore, approaches using TrustZone is more secure than those do not use TrustZone.

These are works on constructing TEE on FPGA-SoC without using ARM TrustZone technology. They focused on augmenting the protection of data and IPs in FPGA-SoC with authentication and integrity checking of bitstream. Kim et al. [21] proposed a framework to secure IPs used for data processing in cloud FPGA-SoC. Wei et al. [23] proposed AccGuard to form TEE for IPs on a cloud FPGA. The TEE was realized via a security manager and remote attestation. However, these schemes did not provide trusted boot and then it is unable to make remote attestation. Therefore, with these schemes, a remote user can not verify trustworthiness of the TEE in the cloud FPGA.

Access control mechanisms were implemented in ShEF [22] via a Shield component to provide secure access to data processed by IPs in the cloud FPGA-SoC. Although ShEF provided a remote attestation process which relied on existing secure boot mechanisms of FPGA, inside attackers were permitted to inject non-secure IPs into the static logic region, which can continuously run on the FPGA to transfer data to or from FPGA TEE. It is vulnerable to IP theft because a software or hardware reconfiguration interface (i.e., PCAP or ICAP) can be utilized by malicious insiders to read out the plain text of IPs in bitstream form configured on FPGA PL.

There are case studies regarding utilizing ARM TrustZone technology to construct TEE on FPGA-SoC. Khan et.al. [17] introduced Open Portable Trusted Execution Environment (OP-TEE), a trusted operating system for ARM TrustZone, to construct FPGA-SoC TEE. But they failed to realize trusted boot (referred to as measured boot proposed in [20]) for remote IP developers. Gross et.al. [20] implemented a firmware Trusted Platform Module utilizing ARM TrustZone and provided trusted boot, where each of the components (including OP-TEE) contained in the trusted boot of FPGA-SoC were measured before being sent to a remote verifier. Moreover, they adopted a reconfiguration interface inside OP-TEE to secure remote bitstream loading, which deploys IPs into PL. However, they do not implement integrity checking of IPs and their FPGA-SoC tbTEE does not support TEE runtime customizing. That is, FPGA-SoC tbTEE is a static TEE. Based on trusted boot and reconfiguration interface in [20], we provide a TEE runtime customizing scheme to deploying IPs in TEE.

Table I summarizes these schemes in terms of security requirements defined in Section 2.5.

TABLE I
COMPARISON OF EXISTING WORKS AND OUR RCTEE †

| Security Require-ments | Not use ARM TrustZone | | | Use ARM TrustZone | | |
|---|---|---|---|---|---|---|
| | [21] 2019 | [23] 2021 | [22] 2022 | [17] 2021 | [20] 2022 | Our RCTEE |
| **SR1** | ○ | ◐ | ● | ○ | ● | ● |
| **SR2** | ● | ● | ● | ○ | ○ | ● |
| **SR3** | - | - | - | ● | ○ | ● |
| **SR4** | ○ | ○ | ○ | ○ | ○ | ● |
| **SR5** | ● | ○ | ● | ○ | - | ● |

† ● full support ◐ partial support ○ no support - not applicable

## 4 SYSTEM DESCRIPTION AND RCTEE OVERVIEW

This section presents the system which uses RCTEE in Section 4.1. Then Section 4.2 and 4.3, respectively, describe the RCTEE architecture and its new components enable the establishment of RCTEE based on FPGA-SoC tbTEE.

### 4.1 System Description

The system includes three participants.

1) Cloud Service Provider (CSP). They earn profit by providing network-attached devices to remote users. They offer a webpage of online store for idle cloud devices with varying PL sizes and hard-wired functionalities.
2) Cloud Service User (USER). These users are the consumers of the CSP. They develop IPs locally and deploy these IPs on a leased cloud device for use.
3) Trusted Third Party (TTP). It plays the role of a trusted authority or is the FPGA vendor, which supports the process of verify TEE on cloud devices as a neutral and trustworthy entity. Their services mainly comprise user enrollment, device enrollment, checking measurements sent by FPGA-TEE and providing users with a FPGA-SoC authentication credential when users request to authenticate the cloud device.

Compared with FPGA-SoC tbTEE, RCTEE is able to support remote attestation, secure IP deployment and IP invocation at runtime. Fig.3 illustrates the phases of customizing RCTEE at runtime (See details of the phases in Section 4.3.4). With the help of TTP, CSP first registers device information and obtains a device-specific bootable image from TTP, and USERs register their information. FPGA-SoC can be booted from the device-specific bootable image to run *T*rusted *O*perating *S*ystem (TOS) of RCTEE in PS. USER can deploy multiple IPs at one time using full bitstream of a hardware design with several IPs integrated. By interacting with RCTEE, USER can verify the trustworthiness of RCTEE, establish a secure communication path with this TEE, deploy IPs and invoke these IPs at runtime.

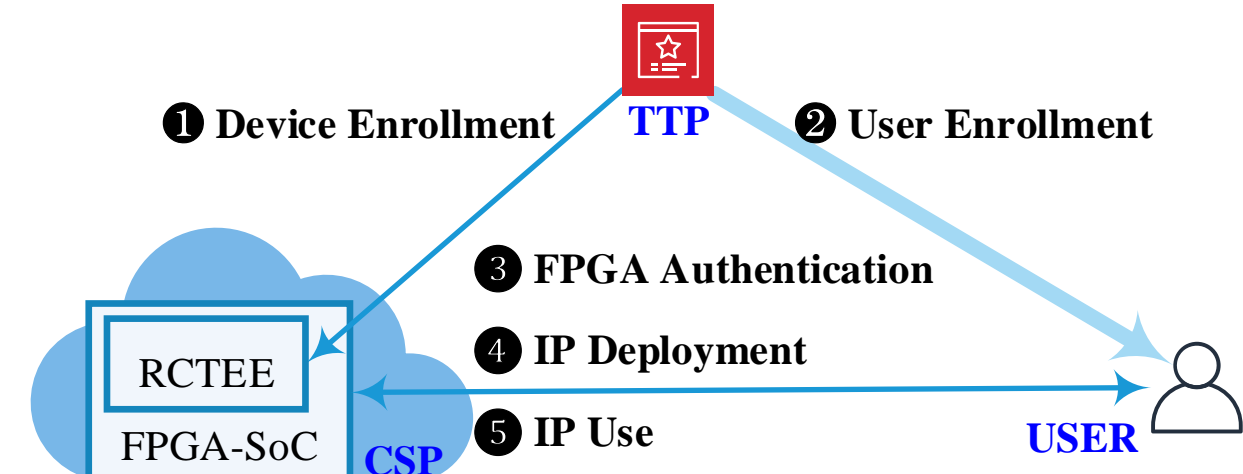

Fig. 3. Illustration of the phases of customizing RCTEE at runtime.

The basic requirements and prerequisites for the cloud device, TTP and the user host are described as follows.

1) **Assumption 1** for the cloud device. We assume a unique device identifier (*#DI*) is programmed by FPGA vendors during manufacturing. The device identifier is non-volatile, unchangeable and readily accessed. However, it is not adequate for authenticating FPGA-SoC in the cloud [11][32].
2) **Assumption 2** for trust on TTP. Several device keys specific to secure boot are programmed in the eFuse or BBRAM of FPGA-SoC by TTP during device enrollment. Note that the challenge-response pairs (CRPs) of PUF are pre-stored in a secure database of TTP for device authentication. An RSA keypair (Public and Private keys, $PK_{TA}$ and $SK_{TA}$) for authenticating trusted applications (TAs) is generated by the TTP. And $PK_{TA}$ is stored in RCTEE to authenticate trusted applications signed with $SK_{TA}$ before running the applications on TOS. Therefore, only TAs signed by TTP can run on RCTEE TOS, which is referred to as TA authentication. CSP operators cannot execute malicious TAs on the TOS unless TAs are signed by TTP, such as the trusted SMA. However, measures are taken to reduce the degree of trust placed in TTP i.e., its access to the encrypted bitstream and data is restricted.
3) **Assumption 3** for protected user host. This work focuses on ensuring the confidentiality and integrity of users' private assets on cloud device. Therefore, we assume that the user host is equipped with a secure execution environment where all critical applications and data are properly protected. All the IPs designed by users are assumed to be secure IPs with two TrustZone specific control port including AWPROT and ARPROT (Section 5.1), which can be protected in proposed scheme.

### 4.2 RCTEE Architecture

RCTEE aims to secure the private data and hardware design consisting of IPs developed by remote users at runtime, suggesting that it must meet **SR1-SR5** given in Section 2.5. Since FPGA-SoC tbTEE can guarantee **SR1**, we need a RCTEE approach which can achieve **SR2-SR5** without damaging **SR1**. We summarize the challenges for **SR2-SR5** as follows:

1) **SR2.** Physically unique and irreproducible fingerprint generator is needed but it does not exist in FPGA-SoC, which makes it difficult to identify FPGA-SoC.
2) **SR3.** There is no reconfiguration interface in OP-TEE but it is needed to deploy IPs of remote users in RCTEE. The reconfiguration interfaces outside of OP-TEE can not be used because they are accessible to inside attackers. In addition, IP protection only with symmetric encryption is not adequate for ensuring the integrity of IPs created by the user. How OP-TEE checks trustworthiness of IPs received from remote users and deploys IP in PL is a big challenge.
3) **SR4.** In the current FPGA-SoC TEE implementations, a new IP needs a new invocation API. That is, when an IP is deployed, the TOS of FPGA-SoC TEE needs modification. How to make a flexible and secure use of any IP deployed at runtime without modifying operating system of FPGA-SoC TEE is a challenge.
4) **SR5.** How to generate a session key for remote users to encrypt/decrypt their data and code, which is unknown to TTP. That is, TTP cannot decode encrypted bitstream and data created by the user.

RCTEE satisfies **SR2-SR5** by developing an extension to FPGA-SoC tbTEE, including three components: CrloadIP, CexecIP and CremoAT, as shown in Fig.4. These components all need to make modifications or implementation in both user space and kernel space of FPGA-SoC tbTEE. Thus, TOS of RCTEE has full access to all components of the device. ROS is exposed to both users and CSP operators. We use a trusted application SMA (secure management application) to denote all the implementation in user space. The modifications to kernel space of TOS include the definition and implementation of new Internal Core APIs (detailed in Section 5) to provide support for secure IP deployment and IP invocation at runtime. Such design aims to maintain a smaller Trusted Computing Base (TCB) size and then avoid the enlarging of attack surface [33][34]. A proxy server is deployed in REE for forwarding data between SMA (that is, CremoAT) and remote users. Therefore, in RCTEE, users can verify the trustworthiness of RCTEE, establish a secure communication path with this TEE, deploy IPs and invoke these IPs at runtime.

We now summarize how RCTEE meets **SR1-SR**5. For **SR1**, we realize a trusted and secure boot process (Section 5.6) based on existing secure boot mechanisms (Section 2.2) and develop CremoAT to verify RCTEE. CremoAT also provides physically unique and irreproducible fingerprint from PUF to authenticate FPGA-SoC for **SR2** and secure communication protocol for **SR5**. For **SR3**, CrloadIP is developed to ensure only RCTEE TOS can deploy IPs in RCTEE when integrity and authentication of bitstream is checked. For **SR4**, CexecIP provides secure and flexible execution of IPs deployed in RCTEE.

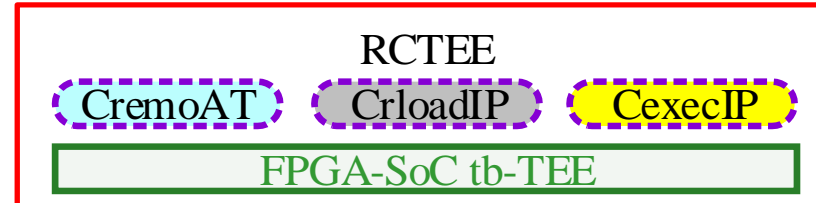

Fig. 4. RCTEE Architecture.

### 4.3 RCTEE Approach

This section first details three components. Then the phases of customizing RCTEE at runtime is presented.

#### 4.3.1 CrloadIP

CrloadIP is comprised of (i) a reconfiguration Internal Core API in kernel space of RCTEE (denoted as CrloadIP-API) and (ii) a functional module named as IP deployment in SMA. In CrloadIP-API, IPs in bitstream form can be configured in FPGA fabric via a software interface (PCAP) or a hardware reconfiguration interface (ICAP). PCAP cannot be isolated from ROS using isolation flow, but this can be achieved by generating custom PMU firmware to restrict both ROS and TOS from accessing PCAP. Consequently, a secure interface for ICAP or PCAP should be integrated in OP-TEE, which allows only TOS to configure PL and restricts insider attacks [17] [20]. Since ICAP occupies parts of hardware resources, users cannot use the entire reconfigurable resources in the cloud FPGA and have to take account the dependence between their hardware design and a static logic with an instance of ICAP. Therefore, we choose PCAP to load bitstream into PL. We develop an Internal Core API (See Section 5.3 for implementation of Reconfiguration Internal Core API) in OP-TEE which is a secure interface for PCAP.

IP deployment in SMA for CrloadIP is responsible for security checking of IPs to be deployed (See Section 5.5 for implementation).

#### 4.3.2 CexecIP

CexecIP consists of two parts: an IP Invocation Internal Core API in kernel space of RCTEE (denoted as CexecIP-API) and a functional module in SMA named as IP invocation (See Section 5.4 for implementation). CexecIP-API is a secure and flexible interface to execute IPs deployed in RCTEE. IP invocation is responsible for processing user request and forwarding input to CexecIP-API for invoking IP. We design a unified input and output format for remote user request, therefore any IP deployed at runtime can be used without modifying operating system of FPGA-SoC TEE.

IP execution needs secure memory. This is achieved by modifying the OP-TEE TOS, namely, reserving part of the physical address space (DDR memory) for input and output of IP invocation when booting FPGA-SoC.

#### 4.3.3 CremoAT

CremoAT is made up with three parts:

(i) A secure PUF IP in PL and an Internal Core API in kernel space of OP-TEE to obtain response from it. We implement this secure PUF IP in initial hardware design and develop an Internal Core API in OP-TEE to obtain FPGA-SoC authentication credential and random seed from it. ARM TrustZone specific control ports are utilized to regulate arbitrary access from ROS to the PUF (See Section 5.1 for implementation).

(ii) An Internal Core API in in kernel space of OP-TEE to read measurements. We modify custom FSBL and OP-TEE to store measurements computed during trusted boot and implement an Internal Core API in OP-TEE to read measurements during run-time (See Section 5.2 for implementation).

(iii) A function module in SMA to perform remote attestation with the cloud service user such that a communication path is established for later secure and trusted data transmission.

#### 4.3.4 Phases of customizing RCTEE

This section details the phases of Fig. 3 as follows. Fig. 5 illustrates the key steps of Phase ❸-❺.

Phase ❶: Device Enrollment

To gain profit from computing power of idle cloud FPGA-SoC, CSP would like to rend out these devices. The first step is to enroll the device information in the database of TTP, and request a device-specific bootable image to support the runtime TEE customizing for FPGA-SoC. TTP reads and stores the unique device identifier (#DI) generated for each device. An RSA keypair ( $SK_{TA}$ and $PK_{TA}$ ) is generated by TTP for authenticating trusted applications (e.g., SMA). Then the SMA is signed with $SK_{TA}$ and the default public key in the source code of OP-TEE is replaced with $PK_{TA}$ . Afterwards, the public key of TTP ( $PK_{TTP}$ ) is stored in SMA for authenticating any certificate of public key signed by TTP. A PUF IP is programmed on the device to collect CRPs. Then, random device keys are generated and programmed into the device. The device keys are used to generate a device-specific bootable image including the following components: ROS with SMA stored in its file system, TOS, firmware and bitstream of an initial hardware design with the PUF IP integrated.

For each device, the TTP creates an entry in a secure device database including the following items: (i) unique device identity (#DI), (ii) CSP's identity (#CSP), (iii) concrete board version of the device, (iv) device keys for secure boot, (v) a keypair for TA authentication ( $SK_{TA}$ and $PK_{TA}$ ) and (vi) a list of CRPs of PUF. Then, the device and bootable image are delivered to the CSP. In our scheme, unique device keys must be generated for each device, which means a unique bootable image needs to be created for each device. When a device is enrolled, CSP publishes the details of all enrolled devices in its online store, and afterwards users can choose the device with specified board version according to their budget. When CSP receives the device and bootable image, the device is booted from the bootable image and SMA is started via the Linux terminal by CSP operators. Then TEE in the cloud FPGA-SoC supporting runtime customizing is ready to receive request from users.

Phase ❷: User Enrollment

Users can register an account for renting cloud devices by sharing their public key ( $PK_{USER}$ ) with TTP. An entry in the database of TTP would be created to record user account (#UID) and $PK_{USER}$ . A user interested in cloud FPGA-SoC can browse through a CSP's online device store and select the ones that meet its requirements. When a user is enrolled, user can receive a certificate of $PK_{USER}$ (i.e., $Ca(PK_{USER})$ ) and $PK_{TTP}$ , and use $Ca(PK_{USER})$ to request the proxy server of SMA.

Phase ❸: Remote Attestation

The remote attestation phrase is a process when TTP verifies measurements for each component during trusted boot of device, then the user authenticates the FPGA-SoC and establish a symmetric session key (SessKey) with SMA. Therefore, they can encrypt the private assets (i.e., bitstream and input data of IPs) before sending it to TEE in the cloud device. ECDH is used to compute SessKey [35]. The protocol flow is illustrated by **step 1-10** in Fig.5. The details are given as follows:

The user makes a request to the proxy server of SMA for remote attestation with $Ca(PK_{USER})$ , **step 1**.

SMA first verify the received $Ca(PK_{USER})$ with pre-stored $PK_{TTP}$ . Only successful verification indicates the user has been registered by TTP and SMA can proceed with the remote attestation process started by the user. Then SMA invokes the PUF IP with pseudo-random input to obtain a random seed and generates an asymmetric keypair for attestation using the seed, a private ( $SK_{DEV}$ ) and public key ( $PK_{DEV}$ ). Then it reads the $H_{BOOT}$ computed in the process of secure boot. SMA concatenates $R(H_{BOOT})$ with device identifier (#DI) and then compute a hash value to obtain $H(H_{BOOT} \| \# DI)$ , abbreviated as α. Next, SMF signs α to obtain a secure boot report $\delta$ . Using $PK_{USER}$ and $SK_{DEV}$ , SMA performs key exchange to compute SessKey. #DI, α and $PK_{DEV}$ is encrypted with $PK_{TTP}$ to obtain a private ε, **step 2**. The secure boot report $\delta$ and private ε are sent to the user, **step 3**.

Once receiving the message from SMA, the user forwards the message to TTP, **step 4**. Combining #DI and $PK_{DEV}$ obtained by decrypting ε with pre-stored CRPs and $H_{BOOT}$ , TTP verifies the secure boot report $\delta$ of the device requested by the user, **step 5**. Only when the requested device is successfully verified, the TTP generates a certificate of $PK_{DEV}$ (i.e., $Ca(PK_{DEV})$ ) and a FPGA-SoC authentication credential. The credential contains a session challenge C, and hash of corresponding response value of PUF concatenated with #DI. TTP sends $Ca(PK_{DEV})$ and the credential to the user, **step 6**.

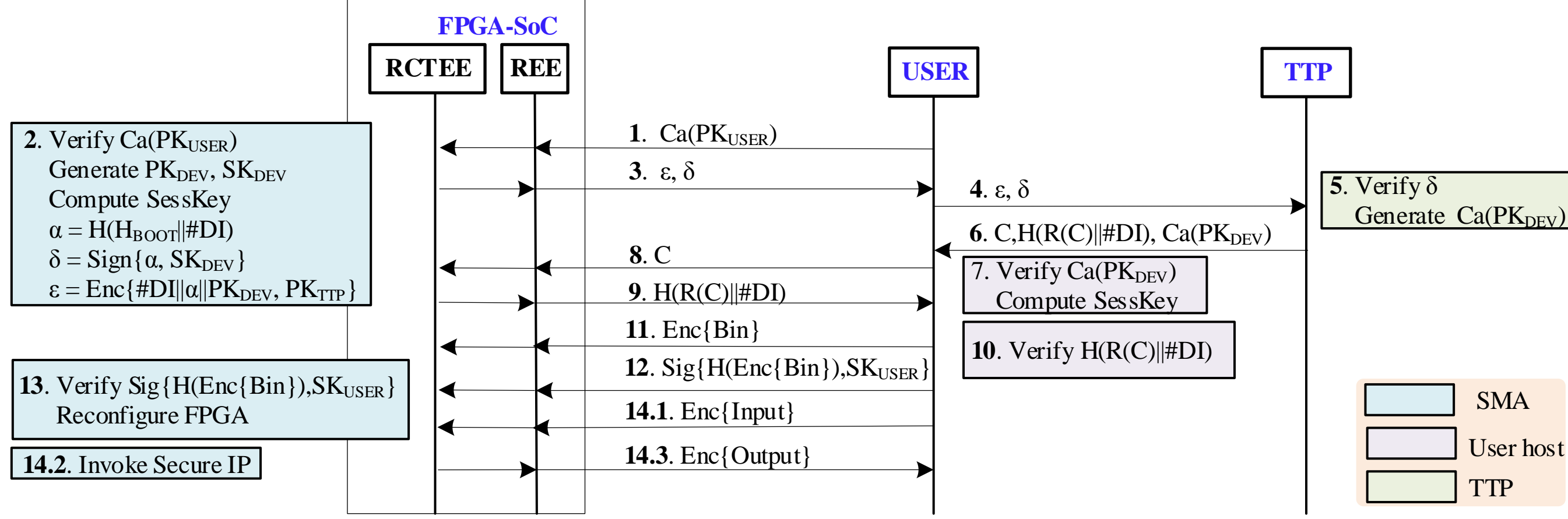

Fig. 5. Main steps in remote attestation, IP deployment, and IP invocation.

After receiving message from TTP, the user can verify $Ca(PK_{DEV})$ with $PK_{TTP}$. Only after a successful validation, the user utilizes $PK_{DEV}$ to compute a SessKey, store the FPGA-SoC authentication credential, and then forwards C to the SMA, **step 7-8**. The SMA invokes the PUF IP to obtain the response of C, computes a $H(R(C)\,||\,\#DI)$, and finally sends it to the user, **step 9**. The user can now authenticate the FPGA-SoC by comparing the value of $H(R(C)\,||\,\#DI)$ received from the TTP and SMA. If the two values are the same, the user successes to verify the cloud device is not impersonated and can use SessKey to communicate with SMA, **step 10**.

Phase ❹: IP Deployment

The main steps of IP Deployment are **step 11-13** presented in Fig.5. The user develops a hardware design integrated with several secure IPs and generates full bitstream, which can be used to deploy IPs in the cloud device. The bitstream is encrypted using SessKey, sent to SMA in RCTEE via the proxy server in ROS [36], **step 11**. The user signs the hash of the encrypted bitstream and obtains $Sig\{H(Enc\{B\text{in}\}), SK_{USER}\}$ .Then the user can request the SMA for reconfiguring the FPGA PL with $Sig\{H(Enc\{B\text{in}\}), SK_{USER}\}$, **step 12**.

SMA generates a hash of the encrypted bitstream file and authenticate it with the $Sig\{H(Enc\{B\text{in}\}), SK_{USER}\}$ received from the user, confirming that the correct bitstream is received. Then SMA reconfigures the FPGA PL with the decrypted bitstream via a secure reconfiguration interface, restricted from ROS, **step 13.**

Phase ❺: IP Invocation

The main steps of IP Invocation are **step 14.1-14.3** presented in Fig.5. The user makes a request to SMA for invoking a secure IP deployed in cloud device with encrypted input data. SMA decrypts the input data and passes the plaintext data to an IP invocation interface. Finally, SMA reads output data via the interface, and send encrypted output data to the user, **step 14**.

## 5 IMPLEMENTATION

In this work, a Xilinx Zynq UltraScale+ XCZU15EG 2FFVB1156 MPSoC is chosen as the target device. We use Xilinx Vitis Design Suite 2020.1, Xilinx Vivado Design Suite 2020.1 and corresponding PetaLinux Tools for the final implementation.

Fig.6 illustrates RCTEE implementation in FPGA-SoC, where sky-blue blocks, grey blocks and yellow blocks represent CrloadIP, CexecIP and CremoAT, respectively. The implementation starts with creating an IP of Ring Oscillator (RO) PUF [37], and exporting the IP as an AXI peripheral with TrustZone specific control ports including AWPROT and ARPROT (Section 5.1). Then a project of the initial hardware design including several initial IPs is built and exported as a XSA file. The initial IPs include an instance of Zynq Ultrascale+ MPSoC, an AXI interconnect IP, and the secure RO-PUF. The XSA file are used in Vivado to create custom firmware (Section 5.2 and Section 5.3). Then we implement four new Internal Core APIs (P1: TEE_GetHwPufResponse, P2: TEE_GetBootHash, P3: TEE_ProgramUserHw, P4: TEE_UsrDefIP) in OP-TEE and apply these APIs to develop trusted functionalities of SMA (Section 5.5). TCP stack provided by Linux are applied to realize untrusted auxiliary functionalities.

### 5.1 PUF-based Random Seed and FPGA-SoC Authentication

The National Institute of Standards and Technology recommends using a truly random number for seeding a Deterministic Random Bit Generator (DRBG) [20]. By doing this with a well-designed DRBG, the output bits produced are unpredictable for an attacker. Although FPGA-SoCs contain a built-in PUF, the noise contained in the PUF response cannot be exploited for obtaining hardware entropy, because this information is not accessible to the developers.

In this work, RO-PUF is adopted to provide random seed for attestation keypair and authenticate FPGA-SoC. A RO-PUF is composed of $N$ identically ROs, two counters, a comparator, and two $N$-bit multiplexers. Every RO oscillates with unique frequency owing to the device's manufacturing process variations. The challenge of the RO-PUF is applied as the select signals of the multiplexers, and thus one pair of ROs is selected. By counting and comparing the number of oscillations within a fixed time interval, a response is generated based on which oscillator from the selected RO pair is faster [37]. We package the RO-PUF as an AXI peripheral with TrustZone specific signals, and an AXI Interconnect IP is used to connect the PS and the RO-PUF. Then, we set the AXI master of AXI Interconnect IP to declare RO-PUF as a secure IP, which is only accessible by TOS. In addition, a syscall (syscall_get_hw_puf_response) and corresponding Internal Core API (TEE_GetHwPufResponse) is implemented in the OP-TEE to read and write the input/output physical address spaces of AXI registers for the RO-PUF. The syscall only allows the request from TAs (e.g., SMA) to invoke the secure RO-PUF.

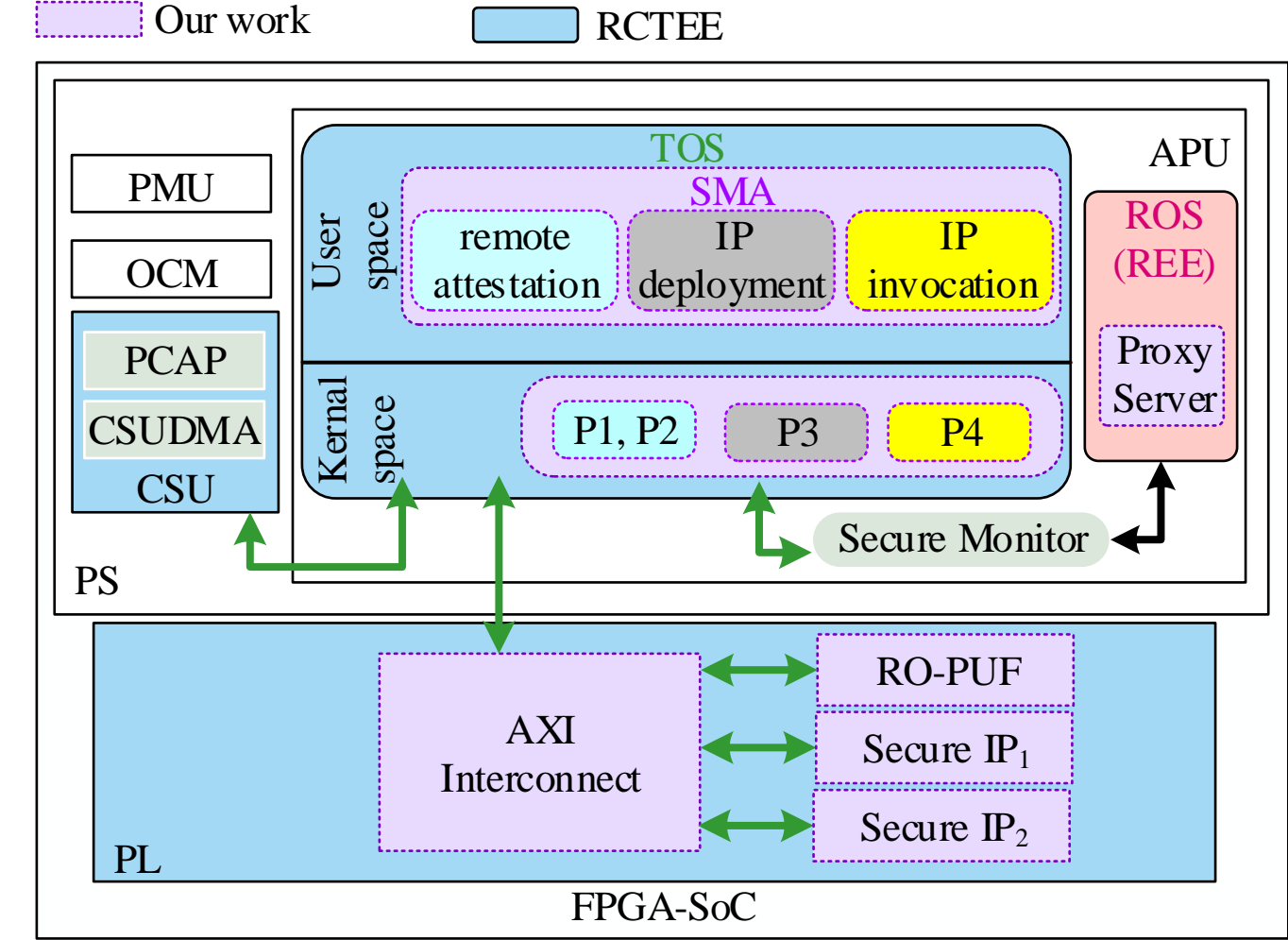

Fig. 6. Illustration of RCTEE implementation on PS and PL in FPGA-SoC.

### 5.2 Storage and Read of Measurement

In comparison to the standard First Stage Boot Loader (FSBL) that is generated from Xilinx tools, we implement a custom FSBL firmware according to the method in [20][38]. With these modifications, the FSBL can measure and check other partitions in bootable image via the SHA3-384 accelerator of CSU, which supports trusted boot. These measurements cannot be directly used for remote attestation, since the FSBL is responsible for loading OP-TEE, which contains the SMA responsible for remote attestation. Therefore, the measurements of the partitions are written in the last chunk of OCM by the custom FSBL before OP-TEE is booted. If the ATF success to initialize OP-TEE, then OP-TEE operating system begins to boot and initialize secure memory region. We modify the boot file (boot.c) in OP-TEE source code to read measurements from the OCM and write these values in secure memory, which can only be accessed by the TOS. Once the measurements are extracted from OCM, the measurements are cleared. When SMA starts to perform attestation with the requested user, it requests a syscall (syscall_get_boot_hash) through a TEE Internal Core API (TEE_GetBootHash) to read these measurements stored in secure memory.

### 5.3 Reconfiguration Internal Core API

Xilinx considers PCAP as trusted under secure boot conditions. However, this interface allows the ROS to load malicious functionality and the readback of IPs in bitstream form [24]. To restrict access to the interface from ROS, we create a custom PMU firmware by removing XilFPGA library form it, and thus PCAP is disabled for both TOS and ROS. To support secure IP deployment through TOS, we modify the OP-TEE source code to include a syscall (syscall_program_user_hw). The syscall only allows the request from TAs (e.g., SMA) to deploy IPs. The syscall transfers bitstreams from secure memory into the PCAP using CSUDMA and invoke the PCAP to program the bitstreams in FPGA fabric. An Internal Core API (TEE_ProgramUserHw) is designed for TAs to requested syscall_program_user_hw in TOS. In order to reconfigure FPGA with a new hardware design including several IPs, users need to add a bitstream setting (i.e., -bin_file) to write the bitstream as a binary bit without header (.bin) in Vivado.

### 5.4 IP Invocation Internal Core API

We design a unified input and output format for IPs and develop an IP invocation Internal Core API (TEE_UsrDefIP) in OP-TEE. For input format, every input data and corresponding input physical address is specified as an input record for IPs. Then, a total number of input records, a physical address for indicating execution state, and a list of output physical addresses for output data should be specified as additional input data. For output format, every output data and corresponding output physical address is binded together as an output record. We reserve secure memory for the input and output physical addresses of IPs and design a syscall (syscall_usr_def_ip) for invoking these IPs through TOS. The syscall first writes the input data of an IP in the corresponding input physical address, then waits for the IP to finish computing by monitoring the execution state, and finally reads output data from the output physical address of the IP. TEE_UsrDefIP is designed for TAs to request the syscall_usr_def_ip in TOS.

### 5.5 Secure Management Application

SMA is implemented as an UMTA [28] encompassing the following functional modules below.

1) Remote attestation module. FPGA-SoC authentication and verifiable TEE is achieved respectively using TEE_GetHwPufResponse and TEE_GetBootHash integrated in TOS. Then an ECDH lib which can generate attestation key pair using the entropy source provided by the RO-PUF is integrated in the SMA. As thus the session key generated through key exchange is sourced from a truly random seed.
2) IP deployment module. Bitstream checking is enforced before configuring FPGA fabric with decrypted bitstream via TEE_ProgramUserHw. Bitstream checking consists of generation of a hash of the encrypted bitstream and verifying it against the provided signed one. Bitstream checking is achieved by utilizing the built-in SHA384 Internal Core APIs of OP-TEE and the integrated EDDH lib.
3) IP invocation module. TEE_UsrDefIP is used to access the secure IPs deployed in FPGA-TEE via requesting the syscall_usr_def_ip in TOS.

### 5.6 Secure and Trusted Boot

eFuse is fixed and non-update-able while a physical battery is required for the BBRAM method [26]. Since eFuse can only be programmed once, which is less flexible than BBRAM, we choose a test mode in this experiment. The test mode is similar to the encrypt only mode (Section 2.2), which replaces eFuse with BBRAM for test and uses custom FSBL firmware to perform trusted boot (Section 5.2). In an actual production process, it is necessary to program eFuse for achieving secure boot mode. In our experiment, integrity-checking of partitions in bootable image is realized by a custom FSBL. Therefore, a 256-bit bitstream encryption key and a 128-bit initialization vector are first generated. Then the key is programmed in the BBRAM of FPGA-SoC.

In the process of packaging, bootable image is created with the following components:

1) Custom FSBL and PMU firmware (PMU_FW) generated by Vitis.
2) Full bitstream (Bitstream) of the PL design generated by Vivado.
3) ARM Trusted Firmware (ATF), ROS (U-BOOT and LINUX) and OP-TEE with Internal Core APIs integrated generated by Petalinux tools.

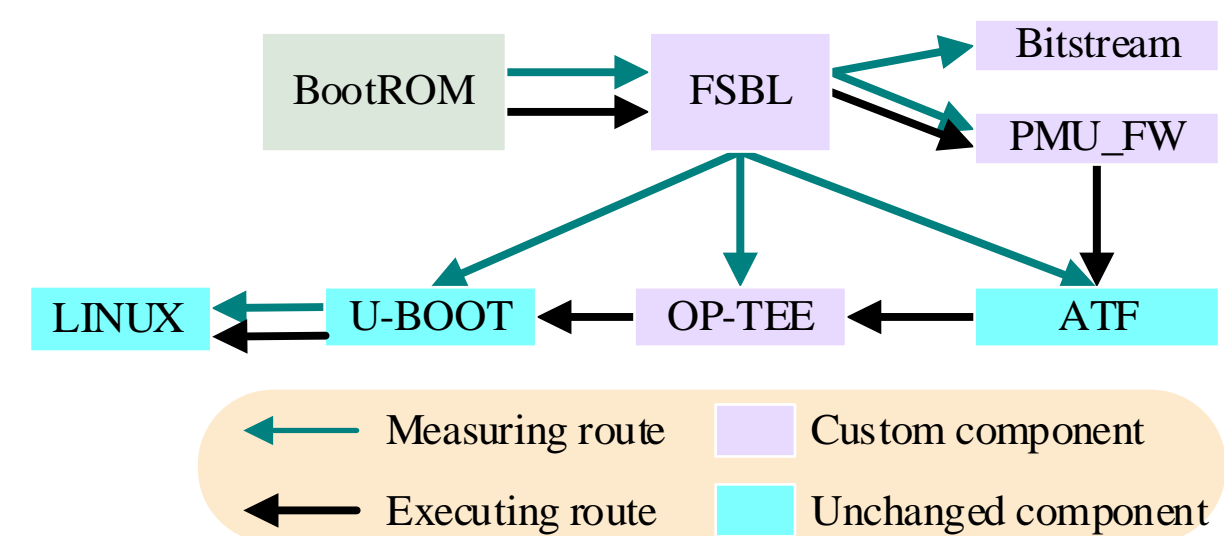

Fig. 7. A secure and trusted boot process with custom components.

These components are encrypted using BBRAM key and packaged into an encrypted bootable image using the bootgen tool [39]. In the process of secure and trusted boot process presented in Fig.7, PMU ROM is first executed upon power on and then it hands over to the CSU ROM. Both PMU ROM and CSU ROM, referred to as BootROM, is fixed and stored on FPGA-SoC in an immutable way, which are the hardware root of trust. CSU decrypts the FSBL before loading it into OCM. The FSBL then performs integrity-checking and decryption before loading the subsequent boot partitions: a full bitstream file (Bit), PMU Firmware (PMU FW), ARM Trusted Firmware (ATF), OP-TEE and u-boot, which loads and measures a Linux Kernel and its root file system [25].

## 6 Security and Performance Analysis

We make security analysis and experimental evaluation in Section 6.1 and 6.2, respectively.

### 6.1 Security Analysis

Table II describes the attacks related to security requirements given in Section 2.5. We analyze the capability of RCTEE against these attacks in the following.

TABLE II
RELATIONSHIP BETWEEN SECURITY REQUIREMENTS AND SECURITY ANALYSIS

| Security Requirements | Cryptographic Attack | Network Attack | | Physical Attack | | |
|---|---|---|---|---|---|---|
| | KPA | RA | MITM | RBA | FIA | UAFR |
| SR1 | | | | √ | √ | √ |
| SR2 | | √ | √ | | | |
| SR3 | | | | √ | √ | |
| SR4 | | | | | | √ |
| SR5 | √ | √ | | | | |

1) Cryptographic Attack

Cryptographic attackers can be an adversary inside or outside the cloud who attempt to break the cryptographic security and obtain the shared session key established between the user and SMA. Security-critical data and bitstream file are protected using encryption when they are in transmission between FPGA-TEE and the user. *Known-plaintext attack* (KPA) is prevented since only ciphertext is known to the attackers. In our designed RCTEE, the plaintext of private assets would never be exposed even if TTP cannot obtain the plaintext shared between FPGA-TEE and the user, preventing KPA attacks.

2) Network Attack

Network attacks are conducted by malicious outsider attackers attempting to impersonate the FPGA-SoC and obtain private assets. The following two types are investigated:

*Replay attack* (RA) is prevented because the session challenge sent by TTP is never repeated, and a random asymmetric key pair for attestation is applied to establish session key. Each session key is used only for the current session and none of the private assets created by previous users can be extracted when a session key is revealed.

*Man-in-the-middle attack* (MITM) is thwarted through utilizing RO-PUF to authenticate FPGA-SoC before the ECDH key exchange. Since the only possible way to access the RO-PUF is restricted to the Internal Core API (`TEE_GetHwPufResponse`), the adversary cannot access the RO PUF through REE.

3) Physical Attack

Physical attackers are assumed to be malicious inside attackers that have the device-specific bootable image and physical access to the cloud FPGA.

*Invasive attack* does not pose any threat to users but to CSP. *Semi-invasive attack* (SIA) can divulge the device keys without damaging the chip. However, attackers have to do multiple reverse engineer attacks on critical components (e.g., OP-TEE and bitstream) in the bootable image, modify them and repackage a new bootable image with a malicious SMA. Furthermore, it's more difficult for attackers to execute the malicious SMA since TA authentication is enforced by the TOS to ensure TA is compiled and signed by TTP.

There exist several possible *Non-invasive attacks. Readback attack* (RBA) aims to read out the bitstream from FPGA PL via a reconfiguration interface even if the bitstream is encrypted before configuration. This attack is prevented via a custom PMU firmware and reconfiguration Internal Core API, which is restricted to SMA. *Fault-injection attack* (FIA) refers to inject malicious IPs to damage the IPs designed by the user or manipulate SMA. It's difficult to program a malicious IP in PL because both authentication and integrity-checking of encrypted bitstream are enforced in SMA. The attack of *Unauthorized access from REE* (UAFR) to secure IPs deployed in FPGA-TEE is blocked because input and output physical addresses of these IPs are protected by ARM TrustZone. Other *Non-invasive physical attacks* cannot be completely prevented but can be mitigated via secure and trusted FPGA-SoC boot. In an actual production process, hardware root of trust mode is desirable for its *side-channel attack* (SCA) resistance. Utilizing hardware cryptographic accelerators together with the secure device keys stored in eFuse or BBRAM, *cold-boot attacks* can be mitigated. In these attacks, an attacker can read private assets from secure memory of TOS by cooling down the memory with a cooling spray such that the data remanence effect lasts longer. An attacker has to rely on a malicious bootloader to read back the contents in secure memory. However, the installation of a malicious bootloader is prevented by authentication and validation provided by the hardware root of trust mode.

### 6.2 Performance Evaluation

In order to evaluate RCTEE performance, we implement a simple hardware design with a LeNet IP integrated, as a design of remote users to customizing FPGA-TEE at runtime. Our experiment is designed and tested on a Xilinx Zynq UltraScale+ XCZU15EG 2FFVB1156 MPSoC. There are mainly six types of available reconfigurable resources on the device:

1) Lookup Tables (LUTs) and Flip-flops (FFs), used for general logic designs.
2) Lookup Table RAMs (LUTRAMs) and Block RAMs (BRAMs), used as memory elements.
3) Global clock buffer (BUFGs), used to drive clock nets in the device.
4) Digital Signal Processing units (DSPs), used to translate signal processing algorithms with multiply-accumulation operations.

We evaluate RCTEE performance in terms *Utilization Ratio* of the above reconfigurable resources and *Time Overhead* of three secure management operations, key exchange, keypair generation and checking of encrypted bitstream.

$$Utilization\ Ratio = Used\ /\ Available$$

Here, *Used* means the amount of reconfigurable resources used by a hardware design and *Available* means the amount of the whole reconfigurable resources on a Xilinx Zynq UltraScale+ XCZU15EG 2FFVB1156 MPSoC. Table III, Table IV and Table V show the results.

Table III shows the resources used by the initial hardware design. Overall, the initial hardware design occupies 4.10% of LUTs, 0.04% of LUTRAMs, 2.94% of FFs, and 0.25% of BUFGs of the FPGA PL. Since the hardware design developed by the user will be programmed to the FPGA PL and overwrite the initial hardware design, the reconfigurable resources occupied by the initial hardware design does not affect the usage of users.

Table IV summarizes the resource utilization ratio of the simple hardware design with a LeNet IP integrated. Different from that in Table III, BRAMs are used to store parameters such as weights and bias that exported from a trained LeNet model. In fact, users can develop personalized hardware design according to their own needs, as long as the usage of various resources does not exceed the amount of available resources. Here we use this simple hardware design to test the feasibility of runtime IP deployment in RCTEE.

TABLE III
RESOURCE USAGE BY THE INITIAL HARDWARE DESIGN

| Re-sources | Used | Avail-able | Utilization Ratio |
|---|---|---|---|
| **LUT** | 18802 | 341280 | 4.10% |
| **LU-TRAM** | 78 | 184320 | 0.04% |
| **FF** | 20069 | 682560 | 2.94% |
| **BRAM** | 0 | 744 | 0.0% |
| **DSP** | 0 | 3528 | 0.0% |
| **BUFG** | 1 | 404 | 0.25% |

TABLE IV
RESOURCE USAGE BY THE HARDWARE DESIGN WITH A LENET IP

| Resources | Used | Available | Utilization Ratio |
|---|---|---|---|
| **LUT** | 14612 | 341280 | 4.28% |
| **LUTRAM** | 78 | 184320 | 0.04% |
| **FF** | 3791 | 682560 | 0.56% |
| **BRAM** | 112.5 | 744 | 15.12% |
| **DSP** | 12 | 3528 | 0.34% |
| **BUFG** | 1 | 404 | 0.25% |

TABLE V
TIME OVERHEAD OF CRITICAL OPERATIONS

| Operation | Time Overhead |
|---|---|
| **Keypair generation** | 5ms |
| **Key exchange** | 2ms |
| **TEE_GetHwPufResponse** | 53ms |
| **TEE_GetBootHash** | 51ms |
| **Checking of encrypted bit-stream** | 0.01345ms/KB |
| **TEE_ProgramUserHw** | 0.0071ms/KB |
| **Receive-Write Bitstream** | 0.2242ms/KB |

We evaluate *Time Overhead* of critical operations of SMA. A bitstream file of 27 MB is generated from the hardware design with a LeNet IP and used to test *Time Overhead*. Table V shows the results, from which we observe:

- **Keypair generation** and **Key exchange**. They are 5ms and 2ms, respectively.
- **TEE_GetHwPufResponse** and **TEE_GetBootHash**. The execution time of these two Internal Core APIs is 53ms and 51ms, respectively.
- **Checking of encrypted bitstream**. The checking on encrypted bitstream, including hashing step, bitstream decryption, and verifying signature, takes 0.01345ms/KB.
- **TEE_ProgramUserHw** and **Read-Write Bitstream**. TEE_ProgramUserHw transfers the full bitstream file to PCAP from secure memory and programmed in PL in 0.0071ms/KB. In our work, the encrypted bitstream file is first transferred to REE via a proxy server running in REE. REE receives the whole file and writes the bitstream into the shared memory between TEE and REE in 0.2242ms/KB. In fact, the encrypted bitstream file can be transferred to secure memory of TEE directly by implementing a proxy server inside TEE, and then it saves the time for REE to transfer bitstream to shared memory. However, it causes a larger TCB size and attack surface compared with our work.

In summary, RCTEE is feasible in terms of overhead on the Xilinx Zynq UltraScale+ XCZU15EG 2FFVB1156 MPSoC.

## 7 CONCLUSIONS AND FUTURE WORK

This work presents a feasible scheme to provide remote users a runtime customizable TEE on a cloud FPGA-SoC. The focus of this work is practicality, security and trustworthiness to the users. We implement a secure RO-PUF (Ring Oscillator Physical Unclonable Function) IP which provides the random seed for key generation and unique response for FPGA-SoC authentication. For runtime TEE customizing and dynamic IP invocation, we implement reconfiguration Internal Core API and IP invoking Internal Core API , which is restricted from unauthorized access. For inside attacker threats, RCTEE can protect dynamic execution of IP deployment and IP invocation from inside attackers by adopting OP-TEE integrated with new Internal Core APIs and custom firmware (such as Platform Management Unit firmware). For untrustworthiness for remote users, FPGA-TEE is certificated via remote validation protocol for FPGA-TEE. For inadequate protection for IPs, RCTEE ensures bitstream checking of encrypted IPs is performed before IP deployment.

In the security analysis, a malicious TTP scenario is also considered besides attackers from inside or outside of the cloud. Countermeasures are proposed to defend against these threats. Most schemes either have drawbacks for lack of trusted boot, or neglect FPGA-SoC authentication or fail to prevent insider attacks. However, RCTEE addresses all of these issues.

This paper has focused on the feasibility of a TrustZone-assisted runtime customizable TEE construction scheme for FPGA-SoC, which supports secure and dynamic IP deployment and IP invocation. One of our next researches is exploring support for multi-tenancy in FPGA-SoC by partitioning PL into regions and placing independent IPs developed by multi-tenants in different regions.